\documentclass{nature_mod}
\usepackage{graphicx}
\usepackage{amsmath}
\usepackage{bbold}
\usepackage{bbm}
\usepackage{braket} 
\usepackage{bm}

\begin{document}

\title{Correlation Driven Transport Asymmetries Through Coupled Spins} 
\author{M. Muenks$^{1}$, P. Jacobson$^{1}$, M. Ternes$^{1,} \footnote{Corresponding Author: m.ternes@fkf.mpg.de}$ \& 
K. Kern$^{1,2}$} 

\date{\today} 
\maketitle

\begin{affiliations}
 \item Max Planck Institute for Solid State Research, Heisenbergstrasse 1, 70569 
Stuttgart, Germany.
 \item Institut de Physique, \'{E}cole 
Polytechnique F\'{e}d\'{e}rale de Lausanne, 1015 Lausanne, Switzerland.
\end{affiliations}

\begin{abstract}

\end{abstract}


\textbf{Correlation is a fundamental statistical measure of order in interacting quantum systems. In solids, electron correlations govern a diverse array of material classes and phenomena such as heavy fermion compounds, Hund’s metals, high-$T_\text{c}$ superconductors, and the Kondo effect \cite{hewson_kondo_1993, sachdev_quantum_2000, lee_doping_2006, weber_strength_2010, georges_strong_2013}.  Spin-spin correlations, notably investigated by Kaufman and Onsager in the 1940’s \cite{kaufman_crystal_1949}, are at the foundation of numerous theoretical models but are challenging to measure experimentally.  Reciprocal space methods can map correlations \cite{schmidt_electronic_2011}, but at the single atom limit new experimental probes are needed.  Here, we determine the correlations between a strongly hybridized spin impurity and its electron bath by varying the coupling to a second magnetic impurity in the junction of a scanning tunneling microscope. Electronic transport through these coupled spins reveals an asymmetry in the differential conductance reminiscent of spin-polarized transport in a magnetic field \cite{loth_spin-polarized_2010}. We show that at zero field, this asymmetry can be controlled by the coupling strength and is directly related to either ferromagnetic (FM) or antiferromagnetic (AFM) spin-spin correlations.}

\newpage

Using the scanning tunneling microscope (STM) as a manipulation tool, it is possible to construct atomically precise magnetic nanostructures and explore the exchange interaction between neighboring spins on surfaces \cite{hirjibehedin_spin_2006, Khajetoorians16, esat_chemically_2016}. For example, the Ruderman-Kittel-Kasuya-Yosida interaction, an oscillatory exchange mechanism, has been observed for spins on magnetically susceptible platinum surfaces and Neel states have been engineered in antiferromagnetically coupled arrays \cite{zhou_strength_2010, loth_bistability_2012}.  Similarly, the global consequences of correlation, such as the superconducting gap or zero bias anomalies due to the Kondo effect, have been found and explored in STM experiments \cite{fischer_scanning_2007, ternes_spectroscopic_2009}.  Competing energy scales, a telltale sign of strongly correlated systems, have recently come under investigation in the two-impurity Kondo problem and the coupling of magnetic molecules to superconducting hosts \cite{bork_tunable_2011, spinelli_exploring_2015, franke_competition_2011}.  Even with these successes, direct measurements of correlation in nanomagnetic systems have proven elusive \cite{burtzlaff_shot_2015}. To directly determine spin-spin correlations, transport experiments through coupled spins, much in the same manner as coupled mesoscopic quantum dots \cite{georges_electronic_1999,van_der_wiel_electron_2002,meden_correlation-induced_2006}, can be performed with the STM.

Here, we use local spectroscopy to study electronic transport through such a coupled spin system. Each metallic lead, tip and sample, harbors an atomic spin system enabling the coupling between the two spins to be smoothly controlled by varying the tip-sample separation. Our coupled spin system is intrinsically asymmetric; the spin bound to the tip is strongly hybridized with Pt while the spin at the surface is decoupled from the underlying Rh metal by an insulating \emph{h}-BN monolayer. The transport characteristics of this junction reveal a distinctive asymmetry in the differential conductance ($dI/dV$), which is a direct result of spin correlations in the tip.  By taking these correlations into account, we can fully describe and model the observed asymmetries within an electronic transport model. We find correlations up to 60\% between the localized spin state on the tip and the itinerant electrons of the metal host.



Figure 1 sketches our experiment, in which we probe a CoH complex on the 
\emph{h}-BN/Rh(111) surface \cite{jacobson_quantum_2015}.
Using vertical atom manipulation \cite{loth_controlling_2010}, we 
functionalize our initially bare tip apex with a Co atom (Fig.~1b, see methods) 
and subsequently probe a second CoH complex (Fig.~1c). For the Co-functionalized tip apex, we observe  
significant changes in the $dI/dV$ spectra when we vary the conductance setpoint, 
$G_s$ (Fig.~1d). Note that magnetic adatoms on Pt surfaces are subject to strong hybridization with the substrate, making it 
difficult to determine the spin state using local spectroscopy \cite{gambardella_giant_2003, mazurenko_renormalized_2010, wiebe_atomic_2011, schweflinghaus_observing_2016}. Therefore, we describe the Co-functionalized tip as a half-integer spin system
that is strongly interacting with the Pt substrate electrons.

For a detailed look at the change of the the $dI/dV$ spectra, we 
incrementally increase $G_s$. Figure~2a shows the result for the 
non-functionalized, that is, a bare Pt tip. The spectra are characteristic for a 
$S = 1$ spin system with magnetic anisotropy and no level degeneracy, as shown 
in our earlier work \cite{jacobson_quantum_2015}. 
We observe step-like increases in the $dI/dV$ signal due to current induced 
transitions from the ground state to the two excited states (Fig.~2b), the energetic position 
of these transitions does not change when $G_s$ is increased by more than an order of magnitude.
However, by employing Co-functionalized tips and increasing $G_s$ 
over a similar range to the bare tip, the step positions shift to higher 
energies and a conductance asymmetry appears at the energetically higher 
(``outer'') step. Two prototypical sets of spectra measured on different CoH 
complexes and with different Co-functionalized tips are shown in Fig.~2c-d. 
Apart from slightly different excitation energies due to the \emph{h}-BN corrugation 
that influences the magnetic anisotropy \cite{herden_lateral_2014, jacobson_quantum_2015}, these two sets vary 
in their $dI/dV$ asymmetry at high $G_s$. The data in Fig.~2c show  
higher $dI/dV$ at positive bias, while the spectra in  
Fig.~2d show the opposite trend with an enhanced $dI/dV$ at negative bias.

To quantify these changes, we determine the step energies and 
the $dI/dV$ asymmetry, $\eta$, of the outer step for different $G_s$ 
(Fig.~3a-f). The asymmetry, $\eta = (h_n - h_p) / (h_n + h_p)$, is defined by 
the intensity of the outer steps at negative, $h_n$, and positive voltages, 
$h_p$\cite{loth_spin-polarized_2010, von_bergmann_spin_2015}. 
Spectra obtained with Co-functionalized tips at high $G_s$ show an evolution of the 
step energies reminiscent of those produced by Zeeman splitting in an 
external magnetic field oriented along the surface 
normal \cite{jacobson_quantum_2015}. Likewise, the asymmetry resembles 
spectra obtained with a spin polarized tip in an external magnetic 
field \cite{loth_spin-polarized_2010, von_bergmann_spin_2015}. However, the  
changes observed here occur in the absence of an external magnetic field and  
as $G_s$ is increased.

To model these results we employ a spin 
Hamiltonian that includes axial, $D$, and 
transverse magnetic anisotropy, $E$, for the $S = 1$ sample 
spin. Similar to earlier 
experiments \cite{jacobson_quantum_2015}, we find easy-axis anisotropy, $D<0$, 
which favors states with  high magnetic moments, $m_z=\ket{\pm1}$. The 
non-negligible $E$ term leads to non-magnetic superpositions \cite{delgado_emergence_2015}: an antisymmetric ground state, $\frac{1}{\sqrt{2}}\left(\ket{-1}-\ket{+1}\right)$, and a symmetric first excited state, 
$\frac{1}{\sqrt{2}}\left(\ket{-1}+\ket{+1}\right)$ (Fig.~2b).
To further account for the functionalized tip, we 
add a term that explicitly describes the direct  
exchange coupling between the spin on the sample, $S_1$, and the tip, $S_2$:
\begin{equation}
\mathcal{H}_0 = D \hat{S}_{1,z}^2 + E (\hat{S}_{1,x}^2 - \hat{S}_{1,y}^2) + 
\vec{J}_{12} \bm{\hat{S}}_1 \cdot \bm{\hat{S}}_2 + \sum_{i=1}^2 g_i 
\mu_{\text{B}} B_z \cdot \hat{S}_{i,z}, 
\end{equation}
where $\bm{\hat{S}}_i=(\hat{S}_{i,x},\hat{S}_{i,y},\hat{S}_{i,z})$ are the 
corresponding spin operators of the $i$-th spin and $\vec{J}_{12}$ is the 
coupling between the two spin systems. The effect of an external magnetic field, $B_z$, 
is accounted for by Zeeman terms that include 
the gyromagnetic factor for each spin, $g_i$, and the Bohr magnetron, 
$\mu_{\text{B}}$.

We approximate the Co-functionalized tip as $S_2 = 1/2$
and diagonalize $\mathcal{H}_0$ 
yielding six eigenstates, $\ket{\psi_k}$, which are 
energetically twofold degenerated at $B_z = 0$~T. Surprisingly, this simple 
model enables us to fit the evolution of the step energies when we assume that 
the coupling between the two spins is 
either Heisenberg-like, $\vec{J}_{12} = ( J_{12}, 
J_{12} , J_{12}$), (Fig.~3b) or Ising-like, $\vec{J}_{12} = ( 0, 
0, J_{12}$), (Fig.~3c) \footnote{ Note, that $S_2 = 3/2$ leads to similar results.}. We find that the direct exchange  coupling, $J_{12}$, is proportional to the conductance, $G_s$, and both are an exponential function of the distance between the two spins, therefore excluding the magnetic dipolar interaction. We describe this orbital overlap with an AFM coupling, $J_{12} > 0$.
The principle evolution of these six eigenstates with $J_{12}$ is shown in Fig.~3g. We observe
that an increase of $J_{12}$ not only leads to higher energies of the 
excited states but also to a clear separation in states with different total 
magnetic moment, $m_z^t=\langle \hat{S}_{1,z} \rangle +\langle \hat{S}_{2,z} \rangle $, similar to spintronic magnetic anisotropy \cite{misiorny_spintronic_2013}.
Additionally, the coupling results in a concomitant polarization, $\langle 
\hat{S}_{1,z} \rangle$, of the $S = 1$ subsystem, 
counteracting the $m_{1,z}=\ket{-1}$ and $\ket{+1}$ superposition of the four 
energetically lowest states. Here, an exchange coupling of $J_{12} = 2$ 
meV is sufficient to polarize the ground and first excited state doublets 
with weights greater than 0.85 (Fig.~3h).

We now continue to describe the electrical transport though the junction by 
employing  a Kondo-like interaction, $\bm{\hat{\sigma}} \cdot 
\bm{\hat{S}}$, between the tunneling electrons 
and the coupled spin system, with $\bm{\hat{\sigma}} = (\hat{\sigma}_x, 
\hat{\sigma}_y, \hat{\sigma}_z) $ as the standard Pauli matrices and 
$\bm{\hat{S}}=\bm{\hat{S}}_1 \otimes \bm{\hat{S}}_2$ as the combined spin
operator of the two spins.
To understand the appearance of the bias
asymmetry at the outer step of the spectra we focus on the transition
from the ground state which has its main weight in  
$\ket{m_1,m_2}=\ket{1,\downarrow}$ to the excited state, $\ket{0,\downarrow}$, 
(solid black arrow in Fig.~3h). During this transition, the spin at the tip 
stays in the $\ket{\downarrow}$ state while the spin on the sample undergoes a 
change of $\Delta m_z = -1$ from $\ket{1}$ to $\ket{0}$. This angular momentum 
has to be provided by the tunneling electron so that the 
process only occurs if the electron changes from $\ket{\downarrow}$ 
to $\ket{\uparrow}$.
As Pt is polarized by magnetic impurities such as Co\cite{gambardella_giant_2003, wiebe_atomic_2011,
 mazurenko_renormalized_2010}, we expect the functionalized tip to
have an imbalance between spin up and spin down electrons.
Assuming an AFM correlation between the state of the tip's spin system and 
the electrons in the tip, leads to a $\ket{\uparrow}$ polarization, while the 
weak coupling\cite{jacobson_quantum_2015} of the sample spin to the host metal does not lead to any 
significant polarization (Fig.~3i). Therefore, for the highlighted transition, 
the conductance will be enhanced at negative bias and suppressed at positive 
bias, in agreement with the data presented in Fig.~2b.

We implement these correlations into our transport model by describing  
the electron bath in the Pt tip by a density 
matrix, $\hat{\varrho}_2$, which is directly 
correlated to the spin state of the attached Co atom: 
\begin{equation}
\hat{\varrho}_2 = 
\begin{pmatrix}
0.5 & 0 \\
0 & 0.5
\end{pmatrix}
+ C 
\sum_{i=x,y,z}
\langle \hat{S}_{i,2} \rangle
\cdot \hat{\sigma}_i.
\end{equation}
The correlation strength, $C$, has been fitted to the evolution of $\eta$ with 
excellent agreement (Fig.~3d-f). We find an
AFM correlation, $C = -0.50 \pm 0.05$, for the dataset with 
positive asymmetry and a FM correlation, $C = 0.35 \pm 0.04$, for the 
set with negative asymmetry. 
To further highlight the validity and quality of our model, we simulate 
$dI/dV$ spectra by accounting scattering up to third order in the matrix 
elements (see methods) by considering additional exchange processes between the localized spin 
on the sample and substrate electrons (Fig.~2e,~f) \cite{ternes_spin_2015}.


To further clarify the coupling, $J_{12}$, between the spin 1 and spin 1/2, we measure a similar system as in Fig.~1c, subject to an external magnetic field, $B_z$ = 5 T (Fig. 4a, b). For weak coupling (small $G_s$) the spectra show the expected Zeeman-shift of the transition energies and a step asymmetry $\eta$ due to field-induced spin-polarization 
in the tip \cite{loth_spin-polarized_2010}. With increasing coupling, these two effects are counteracted by the previously described state polarization and correlation effects. At strong coupling, this 
results in a spectrum that is similar to a bare $S = 1$ spectrum obtained at zero field. In particular we observe that $\eta$ approaches zero, only consistent with AFM coupling, $J_{12} > 0$, between the two spins. FM coupling, $J_{12} < 0$, does not fit the data as it would further increase the asymmetry with $G_s$ (see Fig. 4c).
This measurement, together with the proportionality of $J_{12}$ with $G_s$, allows us to fix the sign of the direct exchange, $J_{12} > 0$, and distinguish between FM and AFM correlations, $C \in [-1,1]$, within the tip electron bath.

In conclusion, we have shown that the correlation between an atomic spin and an electron bath can be determined by coupling it to a second atomic spin in a tunnel junction. AFM direct exchange coupling was consistently found between the spins, but the correlation of the strongly hybridized spin with the electron bath showed either FM or AFM behavior. Here we note, that different Co adatom binding sites on the Pt tip can lead to a different coupling mechanism with the substrate, especially on a Pt microfacet of unknown structure \cite{yayon_bimodal_2006, schweflinghaus_observing_2016}. Additionally, we cannot exclude coupling to other Co atoms in proximity to the apex atom which could also influence the effective correlation to the tip's electron bath \cite{zhou_strength_2010, loth_controlling_2010}.
Unexpectedly, our measurements show that the FM or AFM correlation with the electron bath is related to the direct exchange coupling which shows either Ising (classical) or Heisenberg (quantum) character.  These correlations introduce a measurable transport asymmetry wholly unrelated to static spin polarization and external magnetic fields and might be used as a method to probe correlated electron materials in an inverted tip-sample geometry.

\section*{Methods}

\subsection{Sample Preparation.}
The Rh(111) surface was prepared with multiple Ar$^+$ sputtering cycles and 
annealing up to a temperature of 1100 K. During the final annealing cycle the 
temperature was stabilized at 1080 K and the surface was exposed to borazine 
$(\text{B}_3\text{H}_6\text{N}_3)$ at $1.2 \times 10^6$ mbar for two minutes 
leading to a self-assembled \emph{h}-BN monolayer. Co atoms were then 
evaporated onto the sample surface at a temperature of $\approx 20$~K from a Co 
rod heated by an electron beam. The CoH complexes form during 
the evaporation from residual hydrogen in the vacuum system.

\subsection{Spectroscopy.}
Spectroscopy ($dI/dV$) was measured using an external lock-in amplifier and 
modulating the bias voltage with a sinusoidal of 0.2 mV amplitude and a 
frequency of 689 Hz. 
The conductance setpoint of the tunnel junction ($G_s = I_s / V_s$) is defined by the 
applied bias voltage to the sample, $V_s$, and the setpoint current, $I_s$. 
This conductance setpoint defines the distance between tip and sample and also the coupling strength $J_{12}$. 
We disable the $I_s$ feedback loop in order to take the $dI/dV$ spectrum at a constant distance 
between tip and sample. 
For measurements in magnetic field, an external field of 5 
T was applied along the surface normal. All experiments were performed in 
ultrahigh vacuum ($\approx 10^{-10}$~mbar) and a base temperature of 
1.4~K.

\subsection{Tip Functionalization.}
Bare Pt tips from 25~$\mu$m wire have been functionalized by positioning the 
tip above a CoH complex at a setpoint of $I_s=20$~pA and $V_s=-15$~mV. From 
this setpoint, we decrease the tip-sample separation until a jump in the current 
is observed. The surface area is then scanned to confirm vertical atom 
manipulation. We assume that the hydrogen detaches from the CoH during this 
manipulation. Successful preparation of Co-functionalized tips results in a 
sharper topographic contrast \cite{loth_spin-polarized_2010}.

\subsection{Simulations.}
To simulate the $dI/dV$ spectra we use a perturbative scattering model 
in which spin-flip 
processes up to the 2nd order Born approximation 
are accounted for and which has been previously successfully used on 
different quantum spin systems \cite{jacobson_quantum_2015, 
spinelli_exploring_2015, ternes_spin_2015, Khajetoorians16}. In 
this model the 
transition probability, $W_{i\rightarrow f}$, for an electron to tunnel between 
tip and sample or vice versa and simultaneously changing the quantum state of 
the dimer system between the initial ($i$) and final ($f$) state is given by:
\begin{equation}
 W_{i\rightarrow f}\propto\left( 
|M_{i\rightarrow
f}|^2+J_{1}\rho_1\sum_{m} \left( \frac{M_{i\rightarrow
m}M_{m\rightarrow
f}M_{f\rightarrow i}}{\varepsilon_i-\varepsilon_m} +\mbox{c. c.}\right) \right)
\delta(\varepsilon_i-\varepsilon_f).
\label{equ:W}
\end{equation}
In this expression, $M_{i\rightarrow j} 
=\sum_{i',j'}\sqrt{\lambda_{i'}\lambda_{j'}}\langle 
\sigma_{j'},\psi_j|\hat{\sigma} \cdot \bm{\hat{S}}|\sigma_{i'},
\psi_i\rangle$ is the Kondo-like scattering matrix element for scattering 
from the state vector $\psi_i$ to 
$\psi_j$ of the coupled 
spin Hamiltonian $\mathcal{H}_0$ (equation (1)). The $\sigma_{i',j'}$ 
are the eigenvectors and $\lambda_{i',j'}$ the eigenvalues of the 
density matrices $\varrho$ in tip and sample (equation (2)) for the electrons 
participating in the scattering process \cite{ternes_spin_2015}. Energy 
conservation between initial state energy $\varepsilon_i$ and final state 
energy $\varepsilon_f$ is obeyed by the delta distribution 
$\delta(\varepsilon_i-\varepsilon_f)$ in equation (\ref{equ:W}).
The first term is responsible for the conductance steps observed in the spectra,
 while the second term leads to logarithmic peaks 
at the intermediate energy $\varepsilon_m$ and scales with the 
dimensionless coupling $J_{\rm 1}\rho_1$ between the sample electrons and the 
CoH spin with $J_{\rm 1}$ as the coupling strength and $\rho_1$ as the density 
of states in the sample close to the Fermi energy \cite{ternes_spin_2015}. For 
the systems discussed in figure 2 we found $J_{\rm 1}\rho_1=-0.1\pm0.03$.

\section*{Acknowledgments}
We thank Oleg Brovko, Lihui Zhou, Sebastian Loth, Maciej Misiorny, Philipp Hansmann, and Fabian Pauly for fruitful discussions as well as Gennadii Laskin for his help with the experiment. P.J. acknowledges support from the Alexander von Humboldt Foundation.  M.M. and M.T. acknowledge support from the SFB 767.

\section*{Author Contributions}
M.T. and K.K. conceived the experiments. M.M. and P.J. performed the STM measurements. M.M. and M.T. performed 2nd and 3rd order theory simulations. All authors discussed the results and contributed to the manuscript.

\section*{References}



\newpage
\begin{figure*}[]
\includegraphics[width=1\columnwidth]{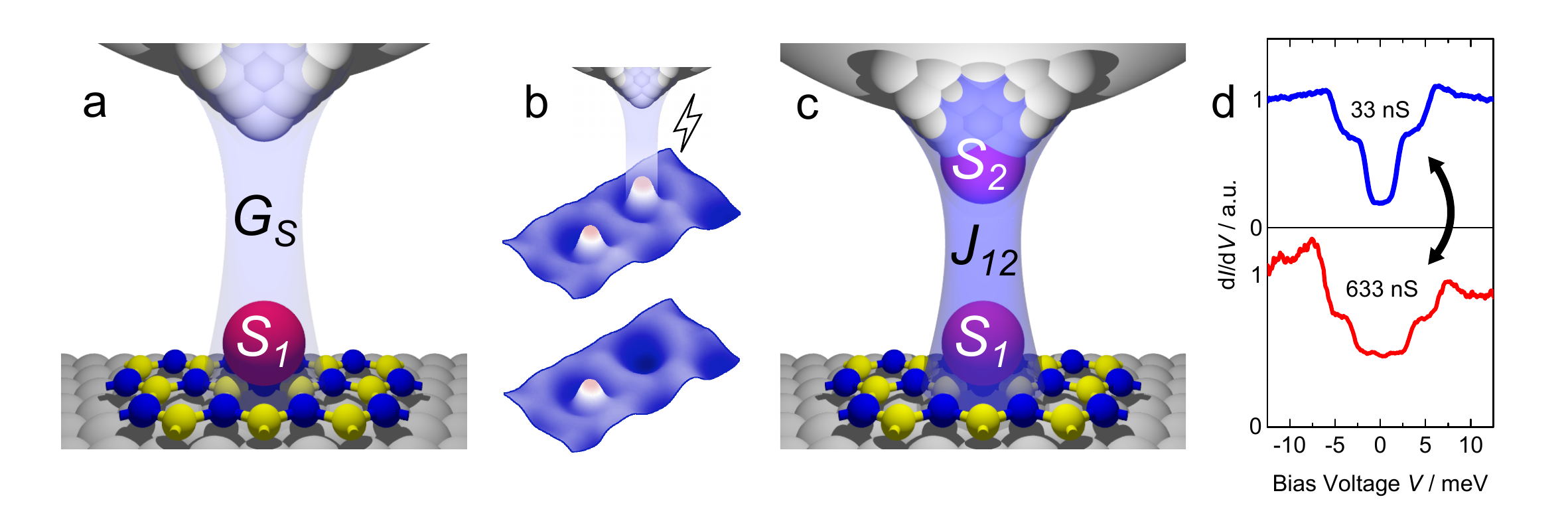}
\caption{Schematics of the experiment. \textbf{a}, The tunnel junction 
consists of two metallic leads, a Pt tip (top) and a Rh sample 
(bottom), harboring a spin, $S_1 = 1$, CoH complex (red sphere) that is 
decoupled by a 
monolayer of \textit{h}-BN (B yellow, N blue spheres). \textbf{b}, 3D 
representations of two successively recorded constant current topographies 
(size $3\times 7$~nm$^2$, $V_s=-200$~mV, $I_s=20$~pA, $T=1.4$~K) illustrating 
the transfer of a Co atom to the tip by vertical atom manipulation. 
\textbf{c}, The tip apex is now a Co atom, strongly hybridized with the Pt tip, 
that we approximate with spin $S_2 = 1/2$. 
\textbf{d}, At low conductance setpoints ($G_s = I_s/V_s= 33$~nS), 
differential conductance ($dI/dV$) spectra obtained with a bare Pt and a 
Co-functionalized tip are identical (blue curve). However, at high $G_s = 
633$~nS the two spins ($S_1$ and $S_2$) are now coupled ($J_{12}$) and the 
spectrum noticeably changes with the Co-functionalized tip (red curve).
}
\label{fig1}
\end{figure*}

\begin{figure*}[]
\includegraphics[width=1\columnwidth]{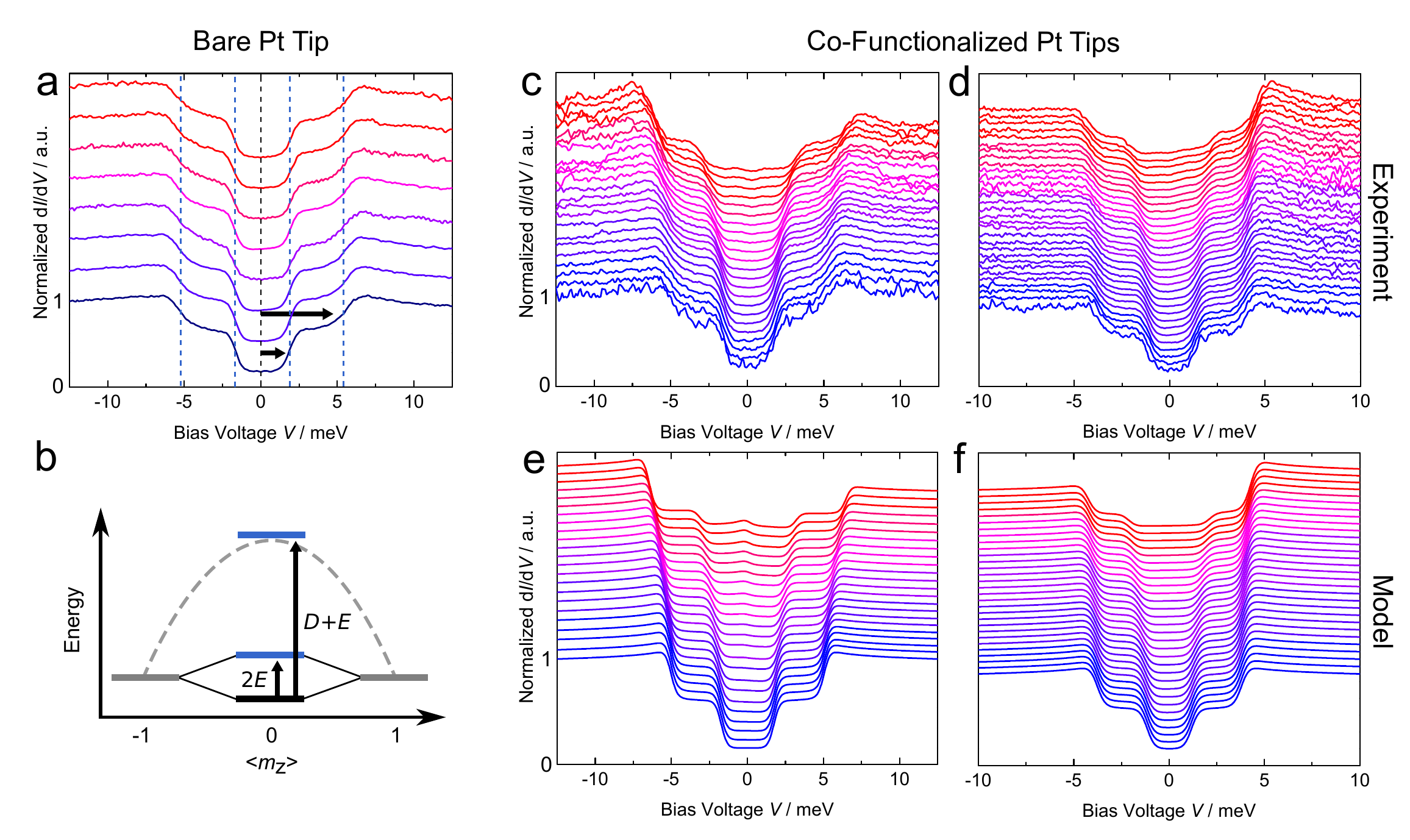}
\caption{Evolution of the $dI/dV$ spectra. \textbf{a}, Spectra obtained 
with a bare Pt tip show no change when increasing $G_s$ from 33 nS (blue) to 
466 
nS (red). The black arrows indicate the two excitation steps at around $\pm 2$ 
meV and $\pm 5$~meV. \textbf{b}, Schematic state diagram of the $S=1$ CoH 
complex in which axial ($D$) and transverse anisotropy ($E$) lift the 
degeneracy. Due to a finite $E$ the magnetic ground and the first excited state 
are superpositions of the $m_z=\ket{+1}$ and $\ket{-1}$ states ($m_z$ is the 
magnetic moment in units of the reduced Planck constant $\hbar$ in the 
$z$-direction, perpendicular to the surface). \textbf{c,d}, Spectra recorded 
with Co-functionalized tips. For low $G_s=50$~nS (blue), both sets are similar 
to the spectra in \textbf{a}. As $G_s$ increases (red), a change of 
the excitation energy occurs and the step at higher energy becomes asymmetric. 
The two different Co-functionalized tips show either at negative or 
positive bias a higher $dI/dV$. \textbf{e}, Simulations 
reproducing the data in (c) with parameters as given in fig.~3 (see methods). 
\textbf{f}, Same as (e) reproducing the 
experimental data in (d).
}
\label{fig2}
\end{figure*}

\begin{figure*}[]
\includegraphics[width=1\columnwidth]{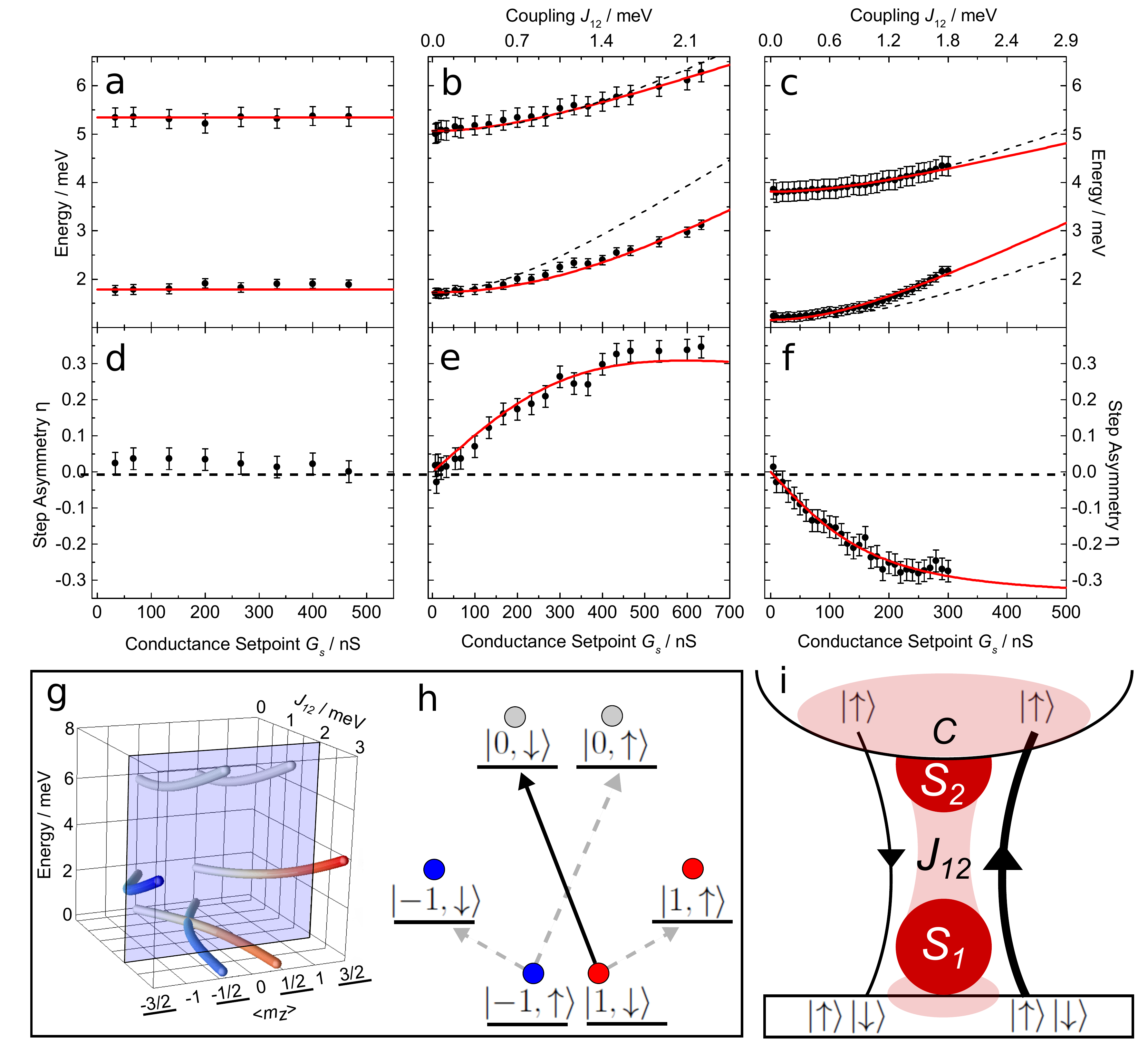}
\label{fig3}
\end{figure*}
\clearpage

\subsection{\rm Figure 3:}
Fits to the transport model. \textbf{a-f}, Evolution of the 
experimentally obtained step energies and asymmetries (black dots) together with
least-square fits to our model (red lines). \textbf{a,d}, For the bare Pt tip, 
the excitation energies remain constant and $\eta\approx 0$. \textbf{b,e}, The 
excitation energies of this set are best described using anisotropy parameters 
of $D = -4.2$~meV, and $E = 0.87$~meV and a Heisenberg-like AFM coupling 
between 
both spins on tip and sample with a strength of $J_{12} = 3.6\ 
\mu\text{eV}/\text{nS}\times G_s$. The coupling to the bath electrons in the 
Co-functionalized tip is accounted for by an AFM correlation strength of $C = 
-0.5 \pm 0.05$. \textbf{c,f}, For this set, an Ising-like AFM coupling between 
both spins with the parameters $J_{12} = 5.9\ \mu\text{eV}/\text{nS}\times 
G_s$ and $D = -3.23$~meV, $E = 0.58$~meV describes the data best. Here the tip 
is FM correlated to the bath electrons with $C = 0.35 \pm 0.04$.	
For comparison an Ising (b) and Heisenberg fit (c) is shown (dashed 
line), not reproducing the experimental data. \textbf{g}, Evolution of the 
state energies and the magnetic moment $m_z$ of the combined system for 
different AFM Heisenberg couplings $J_{12}$ in the two spin system 
($D=-5$~meV, $E=1$~meV). The color code shows the magnetic moment of the 
$S=1$ CoH complex (blue: $-1$, red: $+1$).
\textbf{h}, Cut at $J_{12} = 2$~meV of (g). The main transitions from the 
ground states of the combined spin system are highlighted with arrows (black 
arrow shows the transition sketched in \textbf{i}). 
\textbf{i}, Illustration of the scattering from the state 
$\ket{1, \downarrow}$ to the excited state $\ket{0, \downarrow}$ of the 
combined 
spin system ($S_1 \otimes S_2$) which leads to an asymmetry due to the strongly 
AFM correlated ($C$) electron bath in the tip with the respective $S_2$ 
subsystem.

\begin{figure*}[tp]
\includegraphics[width=1\columnwidth]{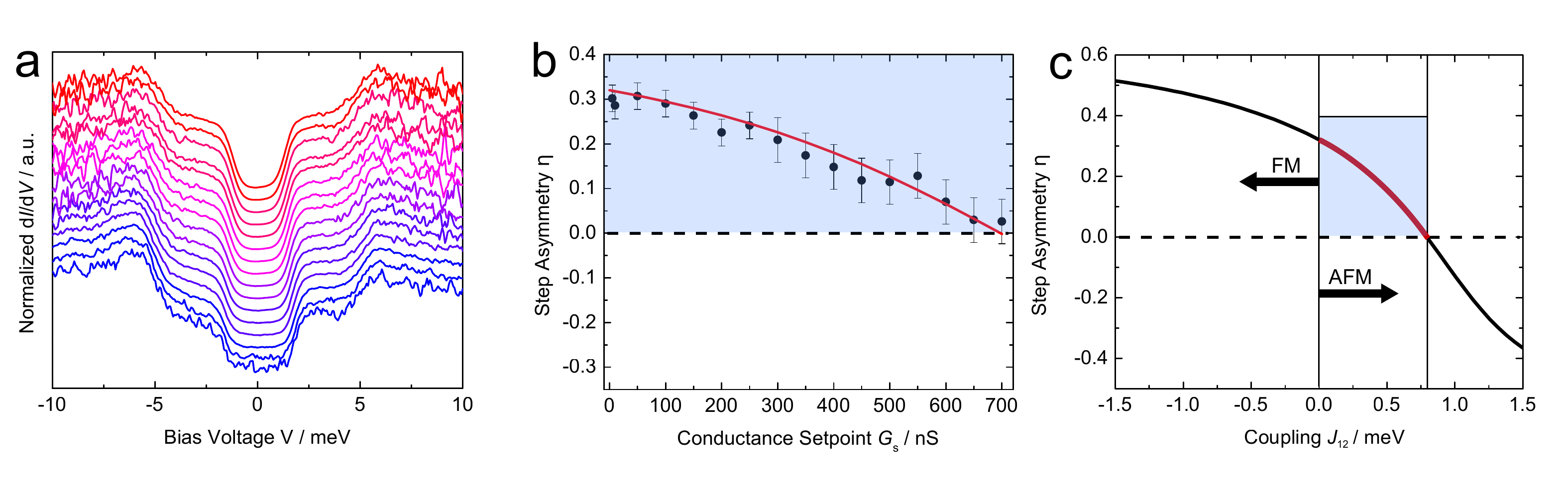}
\caption{Co-functionalized tip spectroscopy in magnetic 
field, $B_z$ = 5 T. \textbf{a}, Evolution of the $dI/dV$ spectra with 
increasing $G_s$ (blue $G_s=20$~nS, red: $G_s=700$~nS). 
\textbf{b}, Fitting the asymmetry, $\eta$, with  anisotropy parameters 
of $D = -4.05$~meV, and $E = 0.65$~meV, yields an Ising-like AFM coupling 
between both spins of $J_{12} = 1.1 \mu\text{eV}/\text{nS}\times G_s$ and 
and a FM correlation of $C = 0.6 \pm 0.05$. At $J_{12}= 0.78$~meV 
the asymmetry, $\eta$, becomes zero. \textbf{c}, A simulation that accounts for 
a wider energy range, shows that only an AFM coupling ($J_{12} > 0$) can 
reproduce the decrease in $\eta$.}
\label{fig4}
\end{figure*}
\clearpage

\end{document}